\documentclass[11pt,prd,superscriptaddress,nofootinbib]{revtex4}
\usepackage[english]{babel}
\usepackage{graphicx}
\usepackage{mathtext}
\usepackage{indentfirst}
\usepackage{epsfig,amsmath,amsfonts}
\usepackage{color}

\newcommand{\beq}[1]{\begin{eqnarray}\label{#1}}
\newcommand\eeq {\end{eqnarray}}
\newcommand\bqa {\begin{eqnarray}}
\newcommand\eqa {\end{eqnarray}}
\newcommand\pr {\partial}

\newcommand\nn {\nonumber}

\newcommand{\bear}{\begin{array}}
\newcommand{\enar}{\end{array}}

\def\th{\mathop{\rm th}\nolimits}
\def\ch{\mathop{\rm ch}\nolimits}

\def\interior#1{\setbox1=\hbox{$#1$}\rlap{$#1$}\kern0.4\wd1\raise1.1\ht1%
\hbox{$\scriptstyle \circ$}}

\def\boxit#1#2{\setbox1=\hbox{\kern#1{#2}\kern#1}%
\dimen1=\ht1 \advance \dimen1 by #1 \dimen2=\dp1 \advance \dimen2 by #1
\setbox1=\hbox{\vrule height\dimen1 depth\dimen2\box1\vrule}%
\setbox1=\vbox{\hrule\box1\hrule}%
\advance \dimen1 by .4pt \ht1=\dimen1 \advance \dimen2 by .4pt \dp1=\dimen2
\box1\relax}
\def\endprf{\raise .5ex\hbox{\boxit{2pt}{\ }}}

\def\ifundefined#1{\expandafter\ifx\csname#1\endcsname\relax}

\def\beq{\begin{equation}}
\def\endq{\end{equation}}
\def\beqa{\begin{eqnarray}}
\def\endqa{\end{eqnarray}}

\renewcommand{\cosh}{\ch}
\renewcommand{\tanh}{\th}

\begin{document}



\centerline{\Large \bf Characters of different secular effects in} \centerline{\Large \bf various patches of de Sitter space}


\vspace{5mm}

\centerline{E. T. ${\rm Akhmedov}^{1, 2}$, U. ${\rm Moschella}^{3}$ and F. K. ${\rm Popov}^{1, 2, 4}$}

\begin{center}
{\it $\phantom{1}^{1}$ B. Cheremushkinskaya, 25, Institute for Theoretical and Experimental Physics, 117218, Moscow, Russia}
\end{center}

\begin{center}
{\it $\phantom{1}^{2}$ Institutskii per, 9, Moscow Institute of Physics and Technology, 141700, Dolgoprudny, Russia}
\end{center}

\begin{center}
{\it $\phantom{1}^{3}$ Universit\`a degli Studi dell'Insubria - Dipartimento DiSAT, Via
Valleggio 11 - 22100 Como - Italy and
INFN, Sez di Milano, Via Celoria 16, 20146, Milano - Italy}
\end{center}

\begin{center}
{\it $\phantom{1}^{4}$ Joseph Henry Laboratories, Princeton University, Princeton, NJ 08544, USA}
\end{center}

\vspace{3mm}

\centerline{\bf Abstract}

There are at least three different types of secular effects in the two--point correlation functions in scalar quantum field theories in de Sitter space--time.
The first one is specific to de Sitter massless and tachyonic minimally coupled scalar fields. The remaining two are generic and are encountered practically in any non--stationary situation in quantum field theory. Furthermore there are secular effects in the  n-point correlation functions  for low enough mass. They are also specific to de Sitter quantum field theory. In this paper we focus on the differences between the secular effects in two--point functions. We discuss also their character in different patches of de Sitter space--time --- global, expanding and contracting Poincar\'{e} patches.

\vspace{10mm}


\section{Introduction}

Secular effects in de Sitter (dS) space quantum field theories have attracted a great deal of attention (see \cite{Starobinsky:1982ee}--\cite{Akhmedov:2013vka}  for an incomplete list of references). Let us consider for instance the two--point Wightman function
\bqa\label{W12}
W(1,2) = \left\langle \phi\left(t_1, \, \vec{x}_1\right) \, \phi\left(t_2, \, \vec{x}_2\right)\right\rangle
\eqa
of a real scalar quantum field on the $D$--dimensional dS manifold (we consider here a  fixed  background, i.e. backreaction effects are not taken into account)\footnote{In this paper we restrict our attention to  real scalar fields, but similar secular effects appear also in other theories.}.
There are at least the following three types of secular effects:

\begin{itemize}
    
\item The {\it secular growth of the first kind} appears in the case of the massless minimally coupled scalars and also for tachyonic fields \cite{Starobinsky:1982ee}--\cite{Tsamis:2005hd} and \cite{bemtach,bemtach2}. It was first seen in the tree--level correlation function (\ref{W12}) when $t_1 = t_2 = t$ and $\vec{x}_1 = \vec{x}_2$ and then observed in the loops in the expanding Poincar\'{e} patch (EPP) $$
ds^2 = dt^2 - e^{2t} \, d\vec{x}^2
$$ of the $D$--dimensional dS space of unit radius. In fact, as $t\to +\infty$ one finds that \cite{Starobinsky:1982ee}--\cite{Tsamis:2005hd}:

\bqa
W_{\rm tree+loops}(1,1) \equiv \left\langle \phi^2\left(t,\vec{x}\right)\right\rangle \approx t \, A_0 + \lambda \, t^3 \, A_1 + \dots
\eqa
$A_0$ is the tree--level contribution,  $A_1$ is the first loop contribution which contains integrals of products of mode functions and $\lambda$ is the self-coupling constant of the scalar field theory.  The dependence on $\vec{x}$ disappears due to the spatial homogeneity of the EPP and the chosen initial state. The effect can also be observed when $\vec{x}_1 \neq \vec{x}_2$ \cite{Gautier:2015pca}--\cite{Moreau:2018lmz}, where the mass term can be treated perturbatively.

This secular growth is specific for massless\footnote{Or to the case when the mass term can be treated perturbatively.} minimally coupled scalars in the EPP and violates dS isometry even at tree--level \cite{Starobinsky:1982ee}--\cite{Tsamis:2005hd}. In fact, in such a case at each order of perturbation theory the Wightman propagator is not a function of the scalar invariant --- the so called hyperbolic distance.  

Methods to deal with  the secular growth of the first kind are developed in \cite{Starobinsky}, \cite{Tsamis:2005hd}, \cite{Gautier:2015pca}--\cite{Moreau:2018lmz}. But they work only in the EPP, for small enough perturbations over the Bunch--Davies (BD) state 
\cite{Chernikov:1968zm}, \cite{Bunch:1978yq}, i.e. only when the same sort of effects in higher point functions can be neglected.

We are not going to discuss effects of this type in detail in the present paper. We have mentioned them just to stress the  difference with respect to the other secular effects on which we are going to focus on in this paper.

\item The {\it secular growth of the second kind} appears for  scalar fields of arbitrary mass; it is seen in the loops when $\left|t_1 -t_2\right| \to \infty$. Namely, in the $\lambda \phi^3$ (resp. $\lambda \phi^4$) theory the one (resp. two) loop corrections to the Wightman propagator contain contributions of the following form:
\bqa
W_{\rm loop}(t_1, \, t_2|p) \approx \lambda^2 \left(t_1 - t_2\right) \, B,
\eqa
where  
\bqa
W(t_1, \, t_2|p) = \int d^{D-1}\vec{x} \, e^{i \, \vec{p} \, \vec{x}} \, \left\langle \phi\left(t_1, \, \vec{x}\right) \, \phi\left(t_2, \, 0\right)\right\rangle
\eqa
is the spatial Fourier transformation of the two--point Wightman function (note that in this paper we always discuss spatially homogeneous states). $W_{\rm loop}$ denotes the one (resp. two) loop contribution, $B$ is a constant containing integrals of products of mode functions whose explicit form will be shown below. The implications of this effect for dS physics have been discussed in the literature (see e.g. \cite{Boyanovsky:1993xf}--\cite{Leblond}). Usually this effect leads to a mass renormalization or to a contribution to the imaginary part of the self--energy. This effect cannot be definitely attributed to the infrared (IR) contributions. It also can appear from the ultraviolet (UV) region of internal loop momenta and even in a atationary situation.

\item The {\it secular growth of the third kind} also appears for  scalar fields of arbitrary mass. It shows up in the loops  when $t_1+t_2 \to \infty$ while  $t_1-t_2 ={\rm const}$. Namely, the one (resp. two) loop corrections to the Wightman function in the $\lambda \phi^3$ (resp. $\lambda \phi^4$) theory in the EPP contain contributions of the form:

\bqa\label{555}
W_{loop}(t_1, \, t_2|p) \approx \lambda^2 \left(t_1 + t_2\right) \, C,
\eqa
where $C$ is a constant containing integrals of products of mode functions. The calculation of the one--loop correction for the Wightman function for the $\lambda \phi^3$ theory with mass $m>(D-1)/2$, in units of dS curvature, has been done in \cite{PolyakovKrotov} both in the EPP and in global dS. The extension of this calculation to the $ \lambda\phi^4$ theory and at higher loops has been done in \cite{AkhmedovKEQ}, \cite{AkhmedovBurda}, \cite{AkhmedovGlobal}, \cite{AkhmedovPopovSlepukhin}, \cite{Akhmedov:2013vka}. The extension to light fields with mass $m<(D-1)/2$ in units of the dS curvature, at one loop was done in \cite{AkhmedovPopovSlepukhin} and at higher loops --- in \cite{Akhmedov:2017ooy}.

\item Finally, in the contracting Poincar\'{e} patch (CPP) and in the global dS manifold there is a {\it secular divergence} in place of the secular growth of the third kind \cite{PolyakovKrotov}, \cite{AkhmedovGlobal}, \cite{Akhmedov:2013vka}:

\bqa
W_{loop}(t_1, \, t_2|p) \approx \lambda^2 \left(t - t_0\right) \, F,
\eqa
where $t=\frac{t_1 + t_2}{2}$, $t_0$ is the initial time (Cauchy surface) from where the self--interactions are adiabatically turned on and $F$ is a constant containing integrals of products of mode functions.   We will see that  the secular growth of the third kind in the EPP  (\ref{555}) and the above secular divergences in the CPP and global dS are of the same physical origin. But, as we explain in the present paper, the resummation of the leading contributions from all loops in the case of the secular growth (of any kind) and in the case of the secular divergence in dS space are physically distinct problems.

The appearance of $t_0$ in the expressions of the correlation functions means the divergence rather than just the growth with time. In fact, if one puts $t_0 \to - \infty$, the loop corrections to the correlation functions are infinite (divergent) even if one cuts off the ultraviolet divergences. It means that in such a case the initial Cauchy surface cannot be taken to past infinity (see \cite{Akhmedov:2013vka} for a generic discussion). This, in its own right, means that in such a situation correlators are not functions of the scalar invariant, i.e. dS isometry is violated in the loops by such secular divergences, even if it is respected at tree--level.

\end{itemize}

Secular effects of the second and third kinds or the secular divergence are generic and appear practically in any non-stationary situation. For example, the secular divergence appears even in flat space quantum field theory in non--stationary situation --- for non--Planckian initial distribution, as discussed in e.g. \cite{LL}. 
Moreover, it  appears in presence of a constant electric field  \cite{Akhmedov:2014hfa,Akhmedov:2014doa,Akhmedov:2009vh}. It is similar to the divergence in global dS and in CPP. 
In the electric pulse in QED there is also a secular growth instead of the divergence, which is similar to the one in the EPP.

The secular growth of the third kind in the case of black hole collapse is discussed in \cite{Akhmedov:2015xwa} (see also \cite{Leahy:1983vb}). At the same time the secular growth of the second kind in the case of black hole collapse was discussed in \cite{Burgess:2018sou}. Finally, the  secular growth of the third kind in the presence of moving mirrors is discussed \cite{Akhmedov:2017hbj} (see also \cite{Astrahantsev:2018lho}).

The secular growth of the third kind and the corresponding divergence in the loops are the infrared effects. These two effects are sensitive to the boundary and initial conditions. As a result it is no wonder that they reveal themselves in different ways in various patches of the same dS space. The presence of a secular growth implies a violation of the applicability of perturbation theory. In fact, even if $\lambda$ is very small $\lambda^2 \, (t_2 - t_1)$, $\lambda^2 \, (t_1 + t_2)$ or $\lambda^2\,\left[(t_1 + t_2)/2 - t_0\right]$ can become of order unity as two arguments $t_{1,2}$ of the correlation function are taken to the future infinity.

Hence, to understand the physics even of massive fields in dS space one has to perform a resummation at least of the leading contributions from all loops. Usually in dS space quantum field theory this is done only for very specific initial states, when the mass term can be treated perturbatively. Meanwhile the result of the resummation strongly depends on the patch and the initial state. The goal of the present paper is to clarify some of these points.

In first place for the resummation of the large infrared effects one has to solve the system of the Dyson--Schwinger equations in some approximation. This system contains equations for the two--point functions and vertexes. Each of these unknowns of the system can possess independent secular contributions. In fact, in $D$--dimensional dS space, when $m\leq \frac{\sqrt{3}}{4}\, (D-1)$ in units of the dS curvature, higher point correlation functions also show a secular growth as one takes all their arguments to the future infinity \cite{Akhmedov:2017ooy} (see also \cite{Bros}--\cite{Bros:2006gs} for the discussion of the origin of such an effect). This phenomenon is also specific to dS space quantum field theory.

In this paper we discuss such situations, when the secular growth in the higher point functions is not present. Namely, we mostly discuss scalars from the principal series, $m>(D-1)/2$. We specifically designate which of the loop contributions provide the leading corrections at each loop level. It happens that in the case of secular growth and in the case of secular divergence  different types of diagrams contribute leading corrections in dS quantum field theory. Hence, e.g. the problems of resummations of leading loop corrections in the EPP and in global dS are physically different.

The paper is organized as follows. In  section II we establish  the setup and the notations. In section III we explain the origin of the secular growth of the third kind in the EPP for the BD state. 
Section IV deals with the difference between the secular growth in the EPP for the BD state and the secular divergences emerging for alpha--vacua in the EPP and for any state in the CPP and in the global dS. 

In  section V we investigate the relation between the secular growth of the second and the third kind for dS invariant situation. In particular, we find the relation between these two types of secular effects for the BD initial state in the EPP in the $x$--space representation.

In section VI we discuss the problem of the resummation of the leading secular contributions from all loops. We explicitly show which type of diagrams provide the leading contributions in the case of secular effects of the second and third kind in the EPP for the initial BD state. Then we show that in the CPP and in global dS different type of diagrams provide the leading contributions in the case of secular divergence.

Section VII contains some conclusions.

\section{Setup}

We consider the scalar field theory:

\bqa\label{freeac}
S = \int d^D x \sqrt{|g|}\, \left[\frac12 \, g^{\alpha\beta} \, \pr_\alpha \phi \, \pr_\beta \phi - \frac12 \, m^2 \, \phi^2 - \frac{\lambda}{3!} \, \phi^3\right],
\eqa
where $\phi$ is real. We restrict our attention to the $\phi^3$ potential just to simplify all expressions. The effects that we discuss here have
nothing to do with the runaway instability of the $\phi^3$ potential and can be seen also in $\phi^4$ theory \cite{AkhmedovPopovSlepukhin}.

The background geometry in (\ref{freeac}) is given by the expanding Poincar\'{e} patch (EPP):

\bqa\label{PP}
ds^2 = \frac{1}{\eta^2}\,\left[d\eta^2 - d\vec{x}^2\right], \quad \eta \equiv e^{-t}.
\eqa
Below we also consider the contracting Poinar\'{e} patch (CPP), global de Sitter (dS) metric and so called sandwich metric --- EPP like expansion interpolating between two flat regions. We set the Hubble constant to one. Only the scalar field $\phi$ is dynamical, gravity is fixed.

The expansion of the field operator over the Bunch--Davies (BD) modes \cite{Bunch:1978yq} is defined as:

\bqa\label{gph}
\phi\left(\eta, \vec{x}\right) = \int \frac{d^{D-1}\vec{p}}{(2\pi)^{D-1}} \, \left[a_{\vec{p}} \, f_{p}(\eta, \vec{x}) + a^+_{\vec{p}} \, f^*_{p}(\eta, \vec{x})\right],
\eqa
where the modes satisfy the following equation:

\bqa
\left[\eta^2 \partial_\eta^2 + (2-d)\eta \partial_\eta + m^2 - \eta^2\Delta \right] f_p\left(\eta, \vec{x}\right) = 0. \label{eoharm}
\eqa
Its solutions are

\bqa
f_p\left(\eta, \vec{x}\right) = \eta^{(D-1)/2} \, h(p\eta) \, e^{- i \, \vec{p}\, \vec{x}}, \quad {\rm and} \quad h(p\eta) = \frac{\sqrt{\pi}}{2} e^{-\frac12 \pi\mu} \, H_{i \, \mu}^{(1)}(p\eta).
\eqa
In the last expression:

\bqa
\mu \equiv \sqrt{m^2 - \left(\frac{D-1}{2}\right)^2}, \quad {\rm and} \quad p \equiv \left|\vec{p}\right|.
\eqa
The normalization factor of the modes fixed by the commutation relations of the field $\phi$ with its conjugate momentum and of the creation, $a^+_{\vec{p}}$, with the annihilation, $a_{\vec{p}}$, operators. 

In this paper we consider the case of $m>0$. Below we discuss the loop IR contributions in the limit, when $p\eta_{1,2} \ll \left|\mu\right|$. In such a limit the leading behaviour of the modes is as follows

\bqa\label{hp1}
h(p\eta) \approx A_+ \, \left(p\eta\right)^{i\mu} + A_- \, \left(p\eta\right)^{-i\mu}, 
\eqa
if $m>(D-1)/2$. Here $A_+ =\frac{2^{-i\mu} e^{-\frac{\pi \mu}{2}} \sqrt{\pi}(1+\coth{\pi\mu})}{2\Gamma(1+i\mu)} \quad {\rm and} \quad A_- = -\frac{i 2^{i\mu} e^{-\frac{\pi \mu}{2}} \Gamma(i \mu)}{2\sqrt{\pi}}$. Most of the equations below are written for the case $m> (D-1)/2$ (for the precise expressions see \cite{Akhmedov:2013vka} for the $m>(D-1)/2$ case and \cite{Akhmedov:2017ooy} for the $m<(D-1)/2$ case). To simplify the discussion below we show just the leading expressions in appropriate limits to reveal the physical meaning of the phenomena that are discussed in the paper. Otherwise expressions become humongous.

In non--stationary quantum field theory every field is described by three propagators (see \cite{LL}, \cite{Kamenev} for an overview). The retarded and advanced propagators are proportional to the commutator, whose spatial Fourier representation at tree--level is equal to:
\bqa\label{C0}
C_0\left(p\left|\eta_1, \, \eta_2\right.\right) \equiv - i \, \int d^{D-1}\vec{x} \, e^{- i \vec{p}\, \vec{x}} \, \left[\phi\left(\eta_1, \vec{x}\right), \, \phi\left(\eta_2, 0\right)\right] = \nn \\ = 2 \, \left(\eta_1 \, \eta_2\right)^{(D-1)/2} \, {\rm Im} \left[h\left(p\eta_1\right) \, h^*\left(p\eta_2\right)\right].  
\eqa
The third relevant two--point correlation function is the Keldysh propagator. Its tree--level spatially Fourier representation is:
\bqa\label{DK0}
D^K_0\left(p\left|\eta_1, \, \eta_2\right.\right) \equiv \frac{1}{2} \, \int d^{D-1}\vec{x} \, e^{- i \vec{p}\, \vec{x}}\left\langle \left\{\phi\left(\eta_1, \vec{x}\right), \, \phi\left(\eta_2, 0\right)\right\}\right\rangle = \nn \\ = \left(\eta_1 \, \eta_2\right)^{(D-1)/2} \, {\rm Re} \left[h\left(p\eta_1\right) \, h^*\left(p\eta_2\right)\right]. 
\eqa
Note that while $C_0$ does not depend on the state, the Keldysh propagator $D^K_0$ does. Eq. (\ref{DK0}) is the expression of the Keldysh propagator for the BD state. 
The index ``$0$'' of $C$ and $D^K$ means that these are the tree--level expressions of the corresponding two--point functions.

Essentially with the use of the Schwinger--Keldysh technique one calculates correlation functions rather than amplitudes, as in the Feynman technique. The result of the calculation solves a Cauchy problem, where the ground state plays the role of the initial state.   Unlike the Feynman technique, the  Schwinger--Keldysh approach is completely causal. The point is that within the Schwinger--Keldysh technique the result of any loop contribution depends only on the causal past of the arguments of correlation functions.

\section{Secular growth of the third kind in the EPP}

Secular effect of the third kind and the corresponding divergence are, in our opinion, potentially the most relevant ones as regards  backreaction on the background dS geometry \cite{Akhmedov:2013vka}, \cite{Akhmedov:2017ooy}. One of the goals of this paper is to show exactly this point.

\begin{figure}[t]
\begin{center} \includegraphics[scale=1]{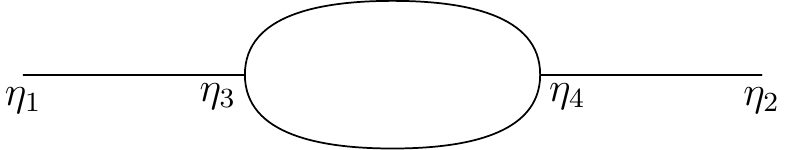}
\caption{In the Schwinger--Keldysh technique there are several diagrams, of the type that is shown here, which contribute to the one loop correction to the two--point functions. The Schwinger--Keldysh diagrammatic technique in the context of cosmology is reviewed in \cite{vanderMeulen:2007ah} (see also \cite{Akhmedov:2013vka}).}\label{1}
\end{center}
\end{figure}

At the leading order the sum of the tree--level and one loop (see fig. \ref{1}) contributions for the Keldysh propagator can be expressed as \cite{Akhmedov:2013vka}:

\bqa\label{EPP}
D_{0+1}^{K}\left(p\, |\eta_1,\eta_2\right) \approx \eta^{D-1}\, \Bigl\{\Bigl[1 + 2 \, n_1(p\eta)\Bigr] \, {\rm Re} \left[h(p\eta_1)\, h^*(p\eta_2)\right] + \Bigl. \nonumber \\ + \Bigr. h(p\eta_1)\, h(p\eta_2) \,\kappa_1(p\eta) +  h^*(p\eta_1)\, h^*(p\eta_2) \,\kappa_1^*(p\eta) \Bigl\},
\eqa
where $\eta = \sqrt{\eta_1\, \eta_2}$ and the modes $h(p\eta_{1,2})$ should be approximated by (\ref{hp1}). In $\phi^3$ theory

\bqa\label{nkappa}
n_1(p\eta) \propto \lambda^2 \, \log\left(\frac{\mu}{p \, \eta}\right)\, \iint_0^\infty \frac{dv}{v} \, dl \, l^{D-2}\, \left[\left|A_+\right|^2 \, v^{i\mu} + \left|A_-\right|^2 \, v^{-i\mu}\right]\, \left[h\left(l/\sqrt{v}\right)\, h^*\left(l\sqrt{v}\right)\right]^2, \nn \\
\kappa_1(p\eta) \propto \lambda^2 \, \log\left(\frac{\mu}{p \, \eta}\right)\, A_+ \, A_- \, \int_1^\infty \frac{dv}{v} \,\int^\infty_{0} dl \, l^{D-2}\, \left[v^{i\mu} + v^{-i\mu}\right]\, \left[h\left(l/\sqrt{v}\right)\, h^*\left(l\sqrt{v}\right)\right]^2.
\eqa
Here $A_\pm$ are defined in (\ref{hp1}) and $\kappa_1^*(p\eta)$ is just the complex conjugate of $\kappa_1(p\eta)$. These expressions are the leading contributions in the limit, when $p\sqrt{\eta_1 \, \eta_2} \to 0$ and $\eta_1/\eta_2 = {\rm const}$. The coefficients of proportionality in (\ref{nkappa}) 
can be found in \cite{Akhmedov:2013vka}. Their exact expression is not necessary for further discussion in this paper. The index 1 in the notation of $n(p\eta)$ and $\kappa(p\eta)$ means that these expressions are just one loop contributions\footnote{In $\lambda \phi^4$ theory the corresponding expressions are similar \cite{AkhmedovPopovSlepukhin} but the secular growth appears in two loop contributions which follow from the sunset diagrams.}. 

We discuss the physical origin of such contributions in the next section. The loop IR corrections to $C_0$ are discussed below in the section V. We will see that in the limit that we are considering in this section $C_0$ does not receive any secularly growing contributions. 

From the form of (\ref{EPP}) it is not hard to recognize in $n(p\eta) \, \left|h(p\eta)\right|^2$ the level population, where $n(p\eta) \, \delta\left(\vec{p} - \vec{q}\right) = \left\langle a^+_{\vec{p}} a_{\vec{q}} (\eta) \right\rangle$,  evaluated at second order in $\lambda$ in the interaction picture. It is important to note that this level population is attributed to the comoving volume. (The volume factor $\eta^{D-1} \equiv \left(\eta_1 \, \eta_2\right)^{\frac{D-1}{2}}$ is the coefficient of proportionality in (\ref{EPP}).) 

Similarly $\kappa(p\eta)\, h^2(p\eta)$ is nothing but the anomalous quantum average $\kappa(p\eta) \,\delta\left(\vec{p} + \vec{q}\right) = \left\langle a_{\vec{p}} a_{\vec{q}} (\eta) \right\rangle$ also evaluated in the interaction picture to the second order in $\lambda$ and attributed to the comoving volume. Finally, $\kappa^*(p\eta)\, \delta^{(3)}\left(\vec{p} + \vec{q}\right) = \left\langle a^+_{\vec{p}} a^+_{\vec{q}} (\eta) \right\rangle$.

Please remember that in this paper we consider spatially homogeneous quantizations only. Also we are working in the interaction picture. Hence, when $\lambda = 0$ all the above quantities, $n$, $\kappa$ and $\kappa^*$, are time independent.
They start to evolve, when one turns on self-interactions.

As $\sqrt{\eta_1 \, \eta_2} \to 0$, i.e. when $(t_1 + t_2)/2 \to + \infty$, we encounter  the secular growth of the third kind. Usually its physical meaning is that due to the $\lambda \, \phi^3$ self--interaction, the  level populations of the low laying exact modes in the theory are changing in time. Also the ground state does change due to the secular growth of the anomalous averages $\kappa_p$ and $\kappa^*_p$ \cite{Akhmedov:2013vka}. That is the usual picture, but in dS space there are peculiarities due to its symmetry. We discuss them in a moment.

In any case, because $\lambda^2 \, \left(t_1 + t_2\right) \gtrsim 1$, for long enough time of evolution, we encounter here a breakdown of  perturbation theory, which is the usual phenomenon in non--stationary situations or even in finite temperature stationary state quantum field theory. This just means that to understand the physics in dS space one has to do at least a resummation of leading secular effects from all loops. The result of the resummation will provide the correct time dependence of $n$, $\kappa$ and $\kappa^*$ rather than just the approximate linear growth.
Consequently, the goal should be to understand which type of contributions are the leading corrections to these quantities at each loop order.

At this point it is important to stress that to perform the resummation of such contributions one usually has to apply the kinetic approach \cite{Kamenev}. However, in dS space there are important peculiarities, which are mainly discussed in the section V. These peculiarities appear because the dS space
has large isometry group which plays the same role as the Poincar\'{e} group in Minkowski space. However, it happens that loop corrections do respect the isometry {\it only} for the exact BD initial state in the exact EPP \cite{Polyakov:2012uc} (see also \cite{Akhmedov:2013vka}).

Meanwhile in the CPP as well as in global dS the isometry group is broken in the loops by IR divergences for any initial state \cite{PolyakovKrotov}, \cite{AkhmedovKEQ}, \cite{AkhmedovGlobal} and \cite{Akhmedov:2013vka}. For alpha--vacua the dS isometry is also broken in the loops even in EPP \cite{Polyakov:2012uc}, \cite{Akhmedov:2013vka}. In the next section we explain the reason for these symmetry violations. 

\section{Secular growth vs. secular IR divergence}

Before discussing the implications of the dS isometry let us consider the origin of (\ref{EPP}) and (\ref{nkappa}) and the situation in the CPP as well as in global dS. For a generic spatially homogeneous background the one loop correction is similar to (\ref{EPP}), but instead of $n(p\eta)$ and $\kappa(p\eta)$ it contains the following expressions:

\bqa\label{basic1}
n^1_p(\eta) \propto \lambda^2 \, \int d^{D-1}q_1 \int d^{D-1}q_2 \, \int_{\eta_0}^\eta d\eta_3 \,\sqrt{g(\eta_3)} \, \int_{\eta_0}^\eta d\eta_4 \, \sqrt{g(\eta_4)} \, \delta\left(\vec{p} + \vec{q}_1 + \vec{q}_2\right) \times \nonumber \\ \times f_p^*\left(\eta_3\right) \, f_p\left(\eta_4\right) \, f_{q_1}^*(\eta_3)\,  f_{q_1}(\eta_4) \, f_{q_2}^*(\eta_3)\, f_{q_2}(\eta_4), \nonumber \\
\kappa^1_p(\eta) \propto \lambda^2 \, \int d^{D-1}q_1 \int d^{D-1}q_2 \, \int_{\eta_0}^\eta d\eta_3 \,\sqrt{g(\eta_3)} \, \int_{\eta_0}^{\eta_3} d\eta_4 \, \sqrt{g(\eta_4)} \, \delta\left(\vec{p} + \vec{q}_1 + \vec{q}_2\right) \times \nonumber \\ \times f_p^*\left(\eta_3\right) \, f^*_p\left(\eta_4\right) \, f_{q_1}^*(\eta_3)\,  f_{q_1}(\eta_4) \, f_{q_2}^*(\eta_3)\, f_{q_2}(\eta_4),
\eqa
and the complex conjugate expression for $\kappa^{1*}_p(\eta)$ \cite{Akhmedov:2013vka}; upper index 1 indicates that we are discussing here one loop corrections. Here $f_p(\eta)$ is the time dependent part of the mode functions, which in the case of the EPP is $f_p(\eta) = \eta^{(D-1)/2}\, h(p\eta)$; $\eta_0$ is the time after which the self--interaction $\lambda$ is adiabatically turned on. In dS space $n_p(\eta) = n(p\eta)$ and $\kappa_p(\eta) = \kappa(p\eta)$ due to the dS isometry invariance, which both in the EPP and CPP contains the simultaneous rescalings of $\eta$ and $\vec{x}$.

When the expressions in Eq. (\ref{basic1}) are not zero, they represent the leading contributions in the limit $\left|\eta-\eta_0\right| \to \infty$, if $\eta$ is the proper time, or in the limit $\eta/\eta_0 \to \infty$, if $\eta$ is the conformal time.

In the flat space case we have that $\sqrt{g(\eta)} = 1$ and $f_p(\eta) = e^{- i \, \omega_p \, \eta}/\sqrt{2\omega_p}$ with $\omega = \sqrt{p^2 + m^2}$ and $\eta$ is the proper (Minkowskian) time. As a result, in such a case in the limit $\left|\eta-\eta_0\right| \to \infty$ one obtains for e.g. $n_p$ the following expression:

\bqa\label{basic}
n^1_p(\eta) \propto \lambda^2 \, \left(\eta - \eta_0\right) \, \int d^{D-1}q_1 \int d^{D-1}q_2 \, \delta\left(\vec{p} + \vec{q}_1 + \vec{q}_2\right) \, \delta\left(\omega_{p} + \omega_{q_1} + \omega_{q_2}\right).
\eqa
Hence, in the situation under consideration the density does not change and remains zero $n_p(\eta) = 0$. Thus, there is no any secular IR divergence of the form $\lambda^2 \, \left(\eta - \eta_0\right)$ due to the energy--momentum conservation:  creation of particles from the ground state is impossible. Similarly $\kappa_p(\eta) = 0$ and the ground state does not change. These are the core facts, which are deeply related to the adiabatic theorem (see e.g. \cite{Trunin:2018egi}--\cite{Leonidov:2018zdi} for the situation in non--stationary systems).

\subsection{Secular growth in the EPP}

Now let us return to the case of the EPP.
For BD modes we have that $f_p(\eta) \sim h(p\eta) \sim e^{i \, p \, \eta}$ when $p\eta \gg \mu$ ($\eta$ is the conformal time). Such a behaviour of the modes in the EPP is the consequence of the strong blue shift of every mode towards past infinity: modes with large physical momenta do not feel the curvature of the space--time and behave as if they were in flat space. Because of that in the expression (\ref{basic1}) the secular growth (\ref{nkappa}) arises only from integrating over momenta and conformal times for which $p\,\eta_{3,4} \ll \mu$ (see \cite{Akhmedov:2013vka} for more details)\footnote{It is important for the remaining part of the paper to understand that to obtain (\ref{nkappa}) from (\ref{basic1}) one has to perform the integration over $q_2$ in (\ref{basic1}) and then make the following change of variables $(q_1, \eta_3, \eta_4) \to (u=\sqrt{\eta_3\, \eta_4}, \, q_1 \, \sqrt{\eta_3 \, \eta_4} = l, \eta_3/\eta_4 = v)$. After that the logarithmic behavior appears as $\int^{\mu/p}_\eta du/u = \log\left(\mu/p\eta\right)$ from the region of integration where $q_1 \gg p$. Note that while expression $(\eta - \eta_0)$ appears in (\ref{basic}) as the consequence of the time translational invariance in Minkowski space, the expression $\log(\mu/p\eta)$ appears in (\ref{nkappa}) due to the conformal time scale invariance of the EPP metric (\ref{PP}). \label{foot1}}.

From these observations one can make two conclusions. First, the limits $\eta_{1,2} \to 0$ and $H\to 0$ do not commute. Here $H$ is the Hubble constant, which is set to one in this note. All the secular growth is gained from the region where mode's physical wave--length exceeds the Compton one, $p\eta < \mu$. That happens due to the space--time expansion, when $H\neq 0$, and mode functions start to behave as is shown in eq. (\ref{hp1}). Summarizing, for the BD state in the EPP we basically have the following situation:

\begin{gather}
n_1(p\eta) \sim \left\{\begin{matrix}
0, \quad p\eta \gg \mu,\\
\lambda^2 \log \frac{p\eta}{\mu},\quad p\eta \ll \mu
\end{matrix}\right.\label{EPP20}
\end{gather}
and similarly for $\kappa_1(p\eta)$ and its complex conjugate.

Second, for the case of the exact BD state in the exact EPP one can take $\eta_0$ to past infinity. In fact, for $p\eta_0 > \mu$ the modes behave as in flat space and one returnes to the situation discussed in the previous subsection. As a result, in (\ref{EPP}) and (\ref{nkappa}) we obtain the secular growth $\log\left(\mu/p\eta\right) \sim (t_1+t_2)/2 - \log\left(p/\mu\right)$ rather than the IR divergence $\log\left(\eta_0/\eta\right) \sim (t_1+t_2)/2 - t_0$.
This fact is crucial for the absence of  secular IR divergence in the case of the exact initial BD state in the exact EPP. Which, in its own right, is important for the dS isometry invariance of the loop integrals in such a situation.

\subsection{Secular IR divergences in various situations in dS space} \label{IVB}

In the case of generic alpha--vacua in the EPP the modes behave as $f_p(\eta) \propto C_+ e^{ip\eta} + C_- e^{-ip\eta}$, when $p\eta \gg \mu$. Here $C_\pm \neq 0$ are complex constants whose values depend on the choice of the alpha--vacuum and one has to plug this expression into Eq. (\ref{basic1}). It follows that  the coefficients of proportionality of $\lambda^2 \, \left(\eta - \eta_0\right)$ are not zero because the arguments of the delta--functions in the corresponding integrals analogous to (\ref{basic}) can be equal to zero. Thus there is an IR divergence, which is to be ascribed to the anomalous UV behaviour of the alpha--modes (for the BD state $C_-=0$, which corresponds to the normal  UV behaviour, i.e. the same as in flat space).

It is probably worth stressing here that $\kappa^1_p(\eta)$ and its complex conjugate also possesses the same secular IR divergence. This means that the system flows to a proper ground state, which is the BD state for $p\eta > \mu$, as one may guess from the proper UV behaviour of the corresponding modes.

In the CPP the situation is as follows. 
Future infinity there corresponds to $\eta \equiv \sqrt{\eta_1\, \eta_2} \to +\infty$ and the BD modes behave as (\ref{hp1}) at past infinity of the CPP\footnote{Here we restrict our attention to the spatially homogeneous states, which are unstable under inhomogeneous perturbations in the CPP, unlike the case of EPP.}. Then, at the leading order, when $p\eta_0 \ll \mu$, the one loop correction to the Keldysh propagator has the same form as (\ref{EPP}), but in the expressions for $n_1(p\eta)$ and $\kappa_1(p\eta)$ in (\ref{nkappa}) one has $\log\left(\mu/p\eta_0\right)$ instead of $\log\left(\mu/p\eta\right)$, if $p\eta > \mu$. At the same time in the case when $p\eta < \mu$ and $\eta/\eta_0 \to \infty$ one obtains $\log\left(\eta/\eta_0\right)$ instead of $\log\left(\mu/p\eta\right)$. 

Summarizing, for the BD initial state in the CPP, when $p\eta_0 \ll \mu$ and $\eta/\eta_0 \to \infty$ we obtain that:

\begin{gather}
n_1(p\eta) \sim \left\{\begin{matrix}
\lambda^2 \log \frac{\eta}{\eta_0}, \quad p\eta \ll \mu,\\
\lambda^2 \log \frac{\mu}{p\eta_0},\quad p\eta \gg \mu,
\end{matrix}\right.\label{CPP21}
\end{gather}
and similarly for $\kappa_1(p\eta)$ and its complex conjugate. Please note the essential difference of this situation from the one in the EPP (\ref{EPP20}). Namely, while in the EPP the evolution of $n(p\eta)$ and $\kappa(p\eta)$'s starts after $\eta\sim \mu/p$, their evolution in the CPP starts right after the initial Cauchy surface $\eta_0$. That is due to the difference of the geometries of the EPP and CPP. 

The coefficients of proportionality in (\ref{CPP21}) are the same as in (\ref{nkappa}), if one considers the BD initial state at the initial Cauchy surface $\eta_0$. Similar secular IR divergences are also present for other alpha--states, but with different coefficients. They are expressed by similar integrals to those in (\ref{nkappa}), but with the corresponding mode functions\footnote{Note that for other alpha--states there also will be secular effects coming from the region $p\eta > \mu$. They are of the same origin as those effects mentioned in the first paragraph of this subsection.}.

In presence of the IR divergence it is impossible to take $\eta_0$ to past infinity (e.g. $\eta_0 \to 0$ in the CPP) because otherwise even after a UV regularization the loop corrections will remain infinite. But keeping $\eta_0$ finite violates the dS isometry, because there are generators of the group that move $\eta_0$. In particular, as a result of that, propagators in $x$--space representation are not functions of the scalar invariant.

In global dS space the situation is similar to the CPP, because it contains EPP and CPP simultaneously. To see the appearance of the IR secular divergence in global dS space one can represent its metric as follows:

\bqa
ds^2 = \frac{1}{\eta^2} \, \left[d\eta^2 - d\vec{x}^2\right], \quad {\rm where} \quad \eta \in \left(-\infty, \, +\infty\right).
\eqa
In such a case the Cauchy surfaces are not compact and the mode functions will be piecewise defined separately for CPP $\eta = e^{t_-} \in [0,+\infty)$ and for EPP $\eta = - e^{-t_+} \in (-\infty, 0]$. Such a situation was considered in \cite{AkhmedovGlobal}, \cite{Akhmedov:2013vka}. Then the IR divergence appears from the CPP part of the loop expressions.

\begin{figure}[ht!]
    \centering
    \includegraphics[scale=1]{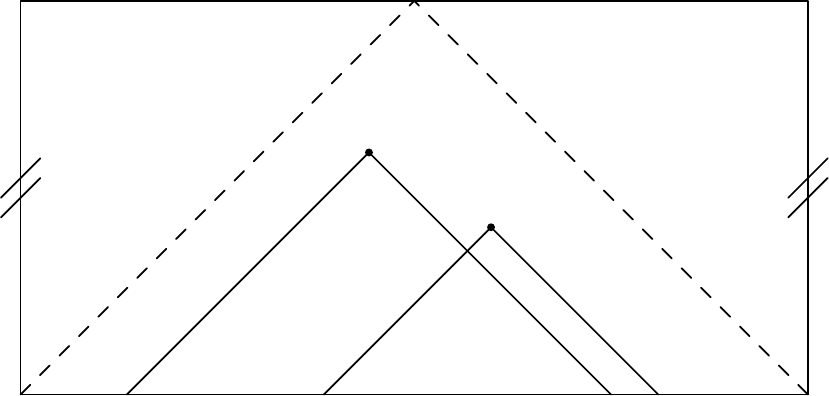}
    \label{fig:PolArg}
    \caption{Here is depicted the Penrose diagram of the 2D dS space. We show that the loop calculation in global dS is similar to the one in the CPP.}\label{11111}
\end{figure}

In fact, the situation in global dS can be understood from the following perspective\footnote{We would like to thank A.Polyakov for communicating to us this argument.}. As we already recalled, the result of a loop calculation within the Schwinger--Keldysh technique depends only on the causal past of the arguments of the correlation function. As can be seen from  Fig. \ref{11111} this essentially means that the result of a loop calculation in global dS  should be the same as in the CPP. If one chooses a different contracting patch from the one shown on this figure, he has to perform a dS isometry transformation, which shifts the patch.

Another option is to consider compact spatial slicing of the global dS space--time:

\bqa
ds^2 = d\tau^2 - \cosh^2(\tau) \, d\Omega^2,
\eqa
where $d\Omega^2$ is the metric on the unit $(D-1)$--dimensional sphere.

To keep the discussion as simple as possible let us explore the 2D global dS space; here the calculations are quite easy to perform and are similar to those in the EPP and CPP. The mode expansion in this case is as follows:

\bqa
\phi\left(\tau, \varphi\right) = \sum_{k=-\infty}^{+\infty} \left[a_k \, f_k(\tau) \, e^{i k \varphi} + a^+_k \, f^*_k(\tau) \, e^{-i k \varphi}\right],
\eqa
where $\varphi$ is the angular coordinate on the spatial circle of the 2D global dS space.

The time dependent part of the modes satisfies the following equation:

\bqa
\left[\partial_{\tau}^2 + \tanh(\tau)\, \partial_{\tau} + \frac{k^2}{\cosh^2(\tau)} + m^2\right] \, f_k(\tau) = 0,
\eqa
which at past and future infinity, as $\tau \to \pm \infty$, becomes similar to the one in the EPP. Indeed, if one makes a change of variables $\tau = -\log \eta$, he approximately recovers  (at the future and past infinity)  Eq. \eqref{eoharm}.  As a result, in this limit the modes behave as:

\bqa
f_k(\tau) \approx \tilde{f}\left(k\, e^{-\tau}\right) \approx \sqrt{k\, e^{-\tau}} \,\left[A_+ \, \left(k\, e^{-\tau}\right)^{-i\mu} + A_- \, \left(k\, e^{-\tau}\right)^{i\mu}\right].
\eqa
We are interested in the so called Euclidean modes, which obey the condition $f_k(-\tau) = f^*_k(\tau)$ and have the normal UV behaviour. These conditions restrict $A_\pm$, but we do not need the corresponding  explicit form. These modes correspond to the BD waves in the EPP. Namely, tree--level two--point correlation functions for these two types of modes coincide with each other.

The leading one loop contribution to the Keldysh propagator in the limit, when initial time is taken to past infinity, $\tau_0 \to -\infty$, and the two arguments of the correlation function are taken to future infinity as  $(\tau_1 + \tau_2)/2 = \tau\to +\infty$, is closely similar to (\ref{EPP}) with:

\bqa
n_k^1(\tau) \propto \lambda^2 \, \sum_{q=-\infty}^{+\infty} \, \int_{\tau_0}^{\tau} d\tau_3 \, \cosh(\tau_3) \, \int_{\tau_0}^{\tau} d\tau_4 \, \cosh(\tau_4) \, f_k^*(\tau_3) \, f_k(\tau_4) \, f_q^*(\tau_3) \, f_q(\tau_4) \, f_{k+q}^*(\tau_3) \, f_{k+q}(\tau_4).
\eqa
Similar expressions hold for $\kappa^1(\tau)$ and its complex conjugate.

The leading contributions to the last expression come from the regions of integration where $\tau_{3,4}\to \tau$, $\tau_{3,4}\to \tau_0$ and $q\gg n$. By changing the variables as

\bqa
u = e^{-(\tau_3 + \tau_4)/2}, \quad l = q \, e^{-(\tau_3 + \tau_4)/2} \quad {\rm and} \quad v = e^{-(\tau_3 - \tau_4)},
\eqa
and replacing of the summation over $q$ with an integral, we get the following expression

\bqa
n^1_k(\tau) \propto \lambda^2 \, \left(\tau - \tau_0 \right) \, \iint \limits_0^{+\infty} \frac{dv}{v} \, dl \, \left[\left|A_+\right|^2 v^{i\mu} + \left|A_-\right|^2 v^{-i\mu}\right] \times \nonumber \\ \times \left[A_+ \, \left(\frac{l}{\sqrt{v}}\right)^{i\mu} + A_- \, \left(\frac{l}{\sqrt{v}}\right)^{-i\mu}\right]^2 \, \left[A_+ \, \left(l\sqrt{v}\right)^{i\mu} + A_- \, \left(l \sqrt{v}\right)^{-i\mu}\right]^2.
\eqa
Here, while the contribution proportional to $\tau$ comes from future infinity, i.e. from that part of global dS which is similar to the EPP, the contribution proportional to $\tau_0$ comes from past infinity, i.e. from that part of global dS which is similar to the CPP.

\section{dS isometry and the relation between the secular growth of the third and the second kind}

As we have mentioned above, in the massive scalar quantum field theory in the EPP there is a dS invariant state, for which the isometry is respected not only at tree--level, but also in all loops \cite{Polyakov:2012uc} (see also \cite{Akhmedov:2013vka}). Such a state is the analogue of the Wightman vacuum in Minkowski space \cite{Bros,bem}. In fact, in the invariant situation nothing depends on the choice of a point in dS space--time. But even in such an invariant situation there is the secular growth of the second and third kind, but there is no secular divergence. Here we discuss the properties  of the secular growth in the $x$--space representation of the correlation functions.

The position space representation of the tree--level BD Wightman function  is as follows:
\bqa\label{W0}
W_0\left(Z_{12}\right) = \left\langle \phi\left(\eta_1, \vec{x}_1\right) \, \phi\left(\eta_2, \vec{x}_2 \right)\right\rangle = \left(\eta_1 \, \eta_2\right)^{\frac{D-1}{2}} \, \int d^{D-1}\vec{p} \, e^{i \, \vec{p} \, \left(\vec{x}_1 - \vec{x}_2\right)} \, h(p\eta_1) \, h^*(p\eta_2) \propto \nonumber \\ \propto \phantom{|}_2F_1\left(\frac{D-1}{2} + i \mu, \frac{D-1}{2} - i \mu; \frac{D}{2}; \frac{1+Z_{12}}{2}\right),
\eqa
where $\phantom{|}_2F_1\left[a, b; c; x\right]$ is the hypergeometric function and $Z_{12} = 1 + \frac{\left(\eta_1 - \eta_2\right)^2 - \left|\vec{x}_1 - \vec{x}_2\right|^2}{2 \, \eta_1 \, \eta_2}$ is the scalar invariant, also called hyperbolic distance between the points 1 and 2. The fact that the correlation functions depend on the scalar invariant reflects the dS  invariance of the state under consideration.

The hypergeometric function in (\ref{W0}) is singular on the light--cone, i.e. when $Z_{12}=1$, and is analytic in the complex $Z_{12}$--plane with the cut going from $Z_{12} = 1$ to infinity along the positive real axis. These values of $Z_{12}$ correspond to time--like separated pairs  of points. To define the correlation function in the vicinity of the cut one has to take the proper boundary value; this is usually encoded in an $\epsilon$ prescription as follows: $W_0\left(Z_{12}\right) \to W_0\left[Z_{12} + i \, \epsilon \, {\rm sign} \left(\eta_2 - \eta_1\right)\right]$. 

Given the  Wightman function one can construct the Keldysh propagator $D^K\left(Z_{12}\right)$, by taking its real part $D^K\left(Z_{12}\right) = {\rm Re}\, W\left[Z_{12} + i \, \epsilon \, {\rm sign} \left(\eta_2 - \eta_1\right)\right]$, and the commutator by taking its imaginary part $C\left(Z_{12}\right) = {\rm Im}\, W\left[Z_{12} + i \, \epsilon \, {\rm sign} \left(\eta_2 - \eta_1\right)\right]$. These relations are true even beyond the tree--level. That is why we drop off the index $0$ in the notations of the Keldysh propagator $D^K$ and the commutator $C$.

Two comments are in order here. The reason why  the dS isometry is respected in the loops for the BD state in the EPP lays in the above analytic properties of the propagator (\ref{W0}) as a function of $Z_{12}$ and in  the specific behaviour of the EPP geometry at past infinity \cite{Polyakov:2012uc} (see also \cite{Akhmedov:2013vka}). These facts are deeply related to the absence of the IR divergences in the loops for the BD ground/initial state in the EPP.

Second, frequently one defines the theory in dS space--time via analytical continuation from the sphere in the complex $Z$--plane (see e.g. \cite{MarolfMorrison}--\cite{Hollands:2010pr}). But such an approach does not allow to address non--vacuum and non--stationary situations in dS cosmology, because in the latter case propagators are not functions of $Z$ anymore. In particular such an approach does not allow one to address the issue of the IR divergences in the CPP and global dS, which are discussed in the previous section.

The limit of interest for us in this note, $p\sqrt{\eta_1\, \eta_2} \to 0$ and $\eta_1/\eta_2 ={\rm const}$, corresponds to the case $Z_{12} \to - \infty$, which is that of the large spatial separation between the points 1 and 2. In such a limit:

\bqa\label{2121}
W_0\left(Z_{12}\right) \approx B_+ Z_{12}^{- \frac{D-1}{2} + i \mu} + B_- Z_{12}^{- \frac{D-1}{2} - i \mu},
\eqa
where $B_{\pm}$ are some complex constants. They can be related, via the inverse Fourier transform of (\ref{C0}), (\ref{DK0}), to the behaviour of the modes in eq. (\ref{hp1}). Otherwise one can obtain eq. (\ref{2121}) from the asymptotics of the hypergeometric function for large values of its argument. 

The one loop correction to the Wightman function in $\phi^3$ theory in the $x$--space representation was calculated in \cite{PolyakovKrotov}. The sum of the tree--level and one loop contributions is as follows:

\bqa\label{W01}
W_{0+1}\left(Z_{12}\right) \approx \left[1 + \lambda^2 \, K \, \log\left(-Z_{12}\right)\right]\, W_0\left(Z_{12}\right), \quad {\rm as} \quad Z_{12}\to - \infty,
\eqa
where $K$ is a constant related to the factors multiplying $\lambda^2 \, \log\left(\mu/p\eta\right)$ in (\ref{nkappa}). Consequently $\log\left(-Z_{12}\right)$ in (\ref{W01}) follows from $\log\left(\mu/p\sqrt{\eta_1 \, \eta_2}\right)$ after the inverse Fourier transformation along the spatial directions. In fact, making the $\epsilon$ shift, $W_{0+1}\left(Z_{12}\right) \to W_{0+1}\left[Z_{12} + i \, \epsilon \, {\rm sign} \left(\eta_2 - \eta_1\right)\right]$, and then taking the real part of the obtained expression one gets the Fourier transformation of (\ref{EPP}) and (\ref{nkappa}) with $h(p\eta_{1,2})$ approximated by (\ref{hp1}) \cite{PolyakovKrotov}. Of course these relations are valid approximately in the limit $p \, \sqrt{\eta_1 \, \eta_2} \to 0$ and $\eta_1/\eta_2 = {\rm const}$.

Taking the imaginary part of $W_{0+1}\left[Z_{12} + i \, \epsilon \,{\rm sign} \left(\eta_2 - \eta_1\right)\right]$ gives that

\bqa\label{C01}
C_{0+1}\left(Z_{12}\right) \approx \left[1 + \lambda^2 \, K \, \log\left|Z_{12}\right|\right]\, C_0\left(Z_{12}\right) + 2 \, \pi \, \lambda^2 \, K \, \theta\left(Z_{12} - 1\right) \, D^K_0 \left(Z_{12}\right).
\eqa
Here $\theta$ is the Heaviside step function. It appears from the imaginary part of the logarithm and should be present here due to the causality properties of the commutator $C$ (see e.g. \cite{Kamenev}). To write it in this form we recall that $\left|Z_{12}\right| \to \infty$.

Eq. (\ref{C01}) exhibits an interesting phenomenon. For space--like separated pairs  ($Z_{12} < 1$) the commutator vanishes  $C_0\left(Z_{12}\right) = 0$, because for these values of $Z_{12}$ the Wightman function $W_0(Z_{12})$ is real. As a result, there is no secular growth in the retarded and advanced propagators in the limit $p \, \sqrt{\eta_1\eta_2} \to 0$ and $\eta_1/\eta_2 = {\rm const}$, i.e. when $Z_{12} \to - \infty$. Thus, only the Keldysh propagator receives secular IR contributions in the limit that we have considered above, which is in agreement with the observations of \cite{Akhmedov:2013vka} and, more generally, of \cite{Kamenev}.

However, for large time--like separations, $Z_{12} \to + \infty$, we have that $C_0\left(Z_{12}\right) \neq 0$ and there is a secular growth in $C_{0+1}\left(Z_{12}\right)$, as follows from (\ref{C01}). This is in agreement with the calculation of the secular loop corrections of  \cite{Leblond}. In the latter paper it was found that in $\phi^3$ theory in the limit $\eta_{1,2}\to 0$ and $\eta_1/\eta_2 \to \infty$ all three propagators (Keldysh, $D^K$, retarded, $D^R$, and advanced, $D^A$) receive the following one--loop contributions:

\bqa\label{D1}
D_1^{K,R,A}\left(p\, |\eta_1,\eta_2\right) \propto \lambda^2 \, \log\left(\frac{\eta_1}{\eta_2}\right) \, D_0^{K,R,A}\left(p\, |\eta_1,\eta_2\right) \, \left[\left|A_+\right|^2 - \left|A_-\right|^2\right] \, \int^{\infty}_1 \frac{dv}{v} \,\int_0^{+\infty} dl \, l^{D-2} \times \nonumber \\ \times \left[v^{i\mu} - v^{-i\mu}\right] \, \left[h\left(l/\sqrt{v}\right) \, h^*\left(l\sqrt{v}\right)\right]^2.
\eqa
This is the secular effect of the second kind, because $\log(\eta_1/\eta_2) = t_2 - t_1$.

Note that $\lambda^2 \, \log\left(\eta_1/\eta_2\right) \approx \lambda^2 \log Z_{12}$ for the time--like $Z_{12}$, when $\eta_1/\eta_2 \to \infty$ and $\vec{x}_1 \approx \vec{x}_2$. Thus, in the dS invariant situation both secular effects of the second and third kinds are related to each other via the isometry group and the analytical continuation in $Z_{12}$.

Below we show that in global dS space and in the CPP the situation with the two secular effects under discussion becomes different. In particular, the problems of resummations of the secular effects and of the secular divergence become physically distinct.

\begin{figure}[t]
\begin{center} \includegraphics[scale=0.8]{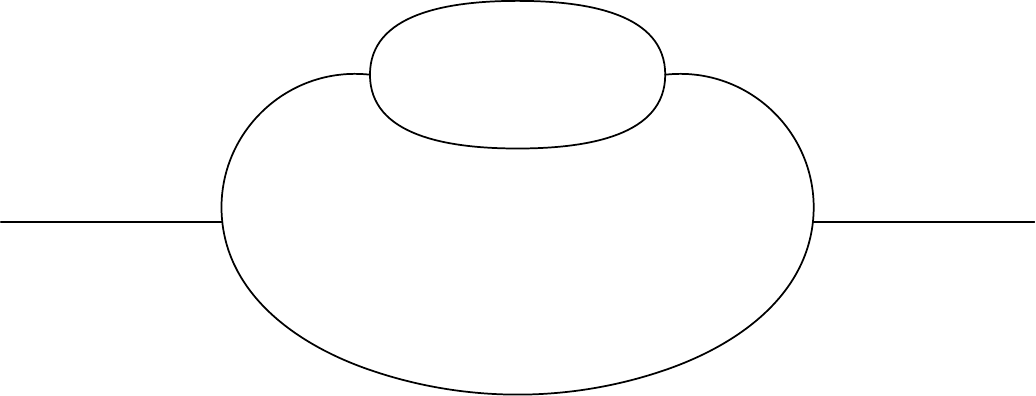}
\caption{In the Schwinger--Keldysh technique there are several diagrams, of the type that is shown here, which contribute to the two loop correction (with bubble inside bubble) to any two--point function.}\label{2}
\end{center}
\end{figure}

\section{Leading vs. subleading higher loop secular corrections}

To perform the resummation of the leading loop secular effects one has to solve the system of Dyson--Schwinger equations. This system is imposed on the two--point functions and on the vertices. As we have mentioned in the Introduction for low enough masses higher--point functions, i.e. vertices, start to grow, when all their arguments are taken to  future infinity \cite{Akhmedov:2017ooy}. In such a situation it is not yet clear how to perform the resummation. Hence, below we restrict ourselves to high enough masses. Our equations are valid for $m>(D-1)/2$. In such a case, if one takes into account  only the leading corrections in powers of $\lambda$ and logarithms, he can put vertices to their tree--level values inside the system of Dyson--Schwinger equations.

In this section we show that for secular effects of the second and third kind only the bubble diagrams of the type shown in Fig. \ref{3} provide the leading contributions in powers of $\lambda^2 \log Z$. At the same time secular effects receive subleading corrections from the diagrams depicted on the Fig. \ref{2}. The latter are suppressed by higher powers of $\lambda$.

On the other hand, in the case of the secular divergence both the diagrams in Fig. \ref{2} and Fig. \ref{3} provide corrections of the same order. As the result, while the resummation of the secular effects is always a linear problem in powers of the exact Keldysh propagator, the resummation of the secular divergences is necessarily non--linear. The last problem has much richer zoo of solutions \cite{Akhmedov:2017ooy}.

\subsection{Exact BD state in the EPP}

Let us start with the correction of the type shown on the Fig. \ref{2}. In such a case in the leading IR corrections one of the propagators in the loops should be represented as (\ref{EPP}) and (\ref{nkappa}). The other propagators should have the tree--level form. This means that instead of $h(q\eta_3) \, h^*(q\eta_4)$ and $h(q\eta_3) \, h(q\eta_4)$, as in (\ref{C0}) and (\ref{DK0}), one loop corrected propagator will contain such contributions as $\lambda^2 \log\left(p\sqrt{\eta_3\eta_4}/\mu\right) \, h(q\eta_3) \, h^*(q\eta_4)$
and $\lambda^2 \log\left(p\sqrt{\eta_3\eta_4}/\mu\right) \, h(q\eta_3) \, h(q\eta_4)$ in the case of third type of secular growth or $\lambda^2 \log\left(\eta_3/\eta_4\right) \, h(q\eta_3) \, h^*(q\eta_4)$ and $\lambda^2 \log\left(\eta_3/\eta_4\right) \, h(q\eta_3) \, h(q\eta_4)$, in the case of the second type of secular growth.

Consider, first, the limit $p\sqrt{\eta_1\, \eta_2} \to 0$ and $\eta_1/\eta_2 = const$, i.e. the third type of secular effect. Then, in this case the second loop of Fig. \ref{2} will contribute to e.g. $n(p\eta)$ the corrections of the following form:

\bqa\label{n2}
n_2(p\eta) \propto \lambda^4 \, \int^{\mu/p}_\eta \frac{du}{u} \, \iint_0^\infty \frac{dv}{v} \, dl \, l^{D-2} \, \log\left(\frac{\mu}{l}\right) \, \left[\left|A_+\right|^2 \, v^{i\mu} + \left|A_-\right|^2 \, v^{-i\mu}\right]\, \left[h\left(l/\sqrt{v}\right)\, h^*\left(l\sqrt{v}\right)\right]^2 \times \nn \\ \times \iint_0^\infty \frac{dv'}{v'} \, dl' \, \left(l'\right)^{D-2} \, \left[\left|A_+\right|^2 \, \left(v'\right)^{i\mu} + \left|A_-\right|^2 \, \left(v'\right)^{-i\mu}\right]\, \left[ h\left(l'/\sqrt{v'}\right)\, h^*\left(l'\sqrt{v'}\right)\right]^2 + \dots.
\eqa
The index 2 here designates that we are discussing second loop corrections, $u$, $l$ and $v$ and their primed versions are defined in the footnote (\ref{foot1}) above; while $v',l'$ are the integration variables corresponding to the internal loop, $v,l$ --- correspond to the big loop in Fig. \ref{2}, and $\log\left(\frac{\mu}{l}\right)$ under the $l$ integral appears from the one loop corrections (\ref{nkappa}). This $dl$ integral is convergent, which is essential for further discussion. 

Ellipses in (\ref{n2}) stand for similar contributions to $n_2(p\eta)$ coming from $\kappa_1(p\eta)$ and its complex conjugate. Their expressions are similar to (\ref{n2}). Moreover, expressions similar to (\ref{n2}) will also appear for $\kappa_2(p\eta)$ and its complex conjugate.

We do not need to know the exact expression for $n_2(p\eta)$ to make the following important conclusion.
The form of Eq. (\ref{n2}) shows that such diagrams as shown in the Fig. \ref{2} (containing loops inside internal propagators) provide contributions of the form $\lambda^4 \log\left(p\, \sqrt{\eta_1 \, \eta_2}\right)$ in the limit under consideration. 

Let us now continue with the consideration of the growth in the limit $\eta_1/\eta_2 \to \infty$, i.e. the second type of secular effect. In such a case one of the internal propagators should have the form (\ref{D1}). As we have discussed at the beginning of this section the leading correction coming from the diagram of the Fig. \ref{2} to all three propagators (R,A and K) will contain contributions as follows:

\bqa\label{D2}
D_2^{K,R,A}\left(p\, |\eta_1,\eta_2\right) \propto \lambda^4 \, \left[D_0^{K,R,A}\left(p\, |\eta_1,\eta_2\right)\right]^2 \, \left[\left|A_+\right|^2 - \left|A_-\right|^2\right]^2  \, \int_{\eta_2}^{\eta_1} \frac{d\eta}{\eta} \times \nn \\ \times \int\limits^{+\infty}_1 \frac{dv}{v} \, \int \limits_0^{+\infty} dl \, l^{D-2} \, \log\left(v\right) \, \left[v^{i\mu} - v^{-i\mu}\right] \,\left[h\left(l/\sqrt{v}\right)\, h^*\left(l\sqrt{v}\right)\right]^2 \times \nn \\ \times \, \int\limits^{+\infty}_1 \frac{dv'}{v'} \, \int\limits_0^{+\infty} dl' \, \left(l'\right)^{D-2} \, \left[\left(v'\right)^{i\mu} - \left(v'\right)^{-i\mu}\right] \,\left[h\left(l'/\sqrt{v'}\right)\, h^*\left(l'\sqrt{v'}\right)\right]^2 + \dots.
\eqa
The $\log(v)$ under the $v$ integral here appears from the first loop correction (\ref{D1}).
Thus, in the limit $\eta_1/\eta_2 \to \infty$ such diagrams as in Fig. \ref{2} contribute $\lambda^4 \log\left(\eta_1/\eta_2\right)$ corrections.

\begin{figure}[t]
\begin{center} \includegraphics[scale=1]{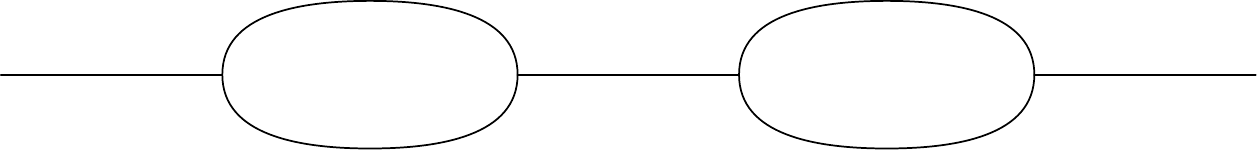}
\caption{In the Schwinger--Keldysh technique there are several diagrams, of the type that is shown here, which contribute multiple bubble type two loop correction to any two--point function.}\label{3}
\end{center}
\end{figure}

At the same time it is quite straightforward exercise to check that the diagrams from the Fig. \ref{3} lead to the contributions of the form $\left[\lambda^2 \log\left(p\, \sqrt{\eta_1 \, \eta_2}\right)\right]^2$ and $\left[\lambda^2 \log\left(\eta_1/\eta_2\right)\right]^2$ in the case of the third and the second kind of secular effects, correspondingly. Thus, if one considers the exact BD state in the exact EPP the diagrams from the Fig. \ref{2} contribute subleading corrections in comparison with those from the Fig. \ref{3}, if $\lambda$ is very small and $\eta_{1,2} \to 0$. This is a very important observation for the resummation procedure.

\subsection{Higher loops in the case of the secular divergence in the CPP and in global dS}

Let us continue now with the discussion of  secular effects in the CPP. The secular growth of the second kind in the CPP has the same properties as in the EPP. The calculations  are practically the same as in the  EPP with the same conclusions that diagrams from  Fig. \ref{2} provide subleading corrections in comparison with  diagrams from  Fig. \ref{3}.

In the case of the secular divergence, which is present instead of the secular growth of the third kind, the situation now is quite different. Because the contribution from the internal loop of the Fig. \ref{2} comes from the past of the external time arguments, $\eta_{1,2}$, the integration over times in this loop are bounded as $\eta_3,\eta_4<u$. As the result the contribution in question has the following form  

\bqa
n_2(p\eta) \propto \lambda^4 \, \int \limits^{\min\left(\eta,\frac{\mu}{p}\right)}_{\eta_0} \frac{du}{u} \, \log\left(\frac{u}{\eta_0}\right) \times \nn \\ \times \left\{\iint_0^\infty \frac{dv}{v} \, dl \, l^{D-2} \, \left[\left|A_+\right|^2 \, v^{i\mu} + \left|A_-\right|^2 \, v^{-i\mu}\right]\, \left[h\left(l/\sqrt{v}\right)\, h^*\left(l\sqrt{v}\right)\right]^2\right\}^2 + \dots.
\eqa
Similar expressions will also appear for $\kappa_2(p\eta)$ and its complex conjugate. 

Thus, the second loop from the Fig. \ref{2} contributes as follows:

\begin{gather}
n_2(p\eta) \sim \left\{\begin{matrix}
\left[\lambda^2 \log \frac{\eta}{\eta_0}\right]^2, \quad p\eta \ll \mu,\\
\left[\lambda^2 \log \frac{\mu}{p\eta_0}\right]^2, \quad p\eta \gg \mu.
\end{matrix}\right. 
\end{gather}
This means that, for the case of the secular divergence the diagrams from the Fig. \ref{2} contribute in the same order as those from the Fig. \ref{3}. That has crucial consequences for the resummation. In particular now the problem of the resummation of the leading IR secular divergences becomes of the kinetic type: one has to derive a dS space analog of the Boltzman's kinetic equation to resum the leading IR divergences \cite{Akhmedov:2013vka}. Meanwhile in the CPP the situation with the resummation of the secular growth of the second kind remains of the same type as in the EPP.

Finally, let us stress that the situation in global dS is again similar to the one in the CPP for the same reason as was explained in the section \ref{IVB}.

\subsection{Perturbations over the BD state in the EPP}

Let us see now what happens if one perturbs the BD state by a non--invariant initial density. In such a case
the initial form of the Keldysh propagator instead of being as in eq. (\ref{DK0}) will be represented by (\ref{EPP}) with $\kappa = 0$ (to have the proper Hadamard behaviour) and some initial distribution $n^0_p$. The retarded and advanced propagators do not depend on the state at the tree--level.

Please recall at this point that $n^0_p$ is the comoving density. Hence, one cannot just put $n^0_p$ at past infinity of the EPP, because then the initial physical density will be infinite. To overcome this problem one has to consider an initial Cauchy surface $+\infty > \eta_0 > 0$ and impose $n^0_p$ there. Let us stress that, if one keeps $n_p^0$ finite, then $\eta_0$ cannot be taken to the past infinity. In this sense now the situation in EPP becomes very similar to the one in the CPP and global dS \cite{Akhmedov:2013vka}. Furthermore, the $x$--space representation of the tree--level Kledysh propagator will not be a function of $Z_{12}$ anymore. Hence, the dS isometry will be broken by the initial condition.

However, despite the presence of the IR cutoff $\eta_0$, the situation for the secular effect of the second kind does not change substantially.
Namely, from the diagram of the Fig. \ref{2} it still has the form (\ref{D1}) and (\ref{D2}) with different coefficients multiplying $\lambda^2 \, \log\left(\eta_1/\eta_2\right)$ and $\lambda^4 \, \log\left(\eta_1/\eta_2\right)$. I.e. for the secular effect of the second kind diagrams from the Fig. \ref{2} still provide subleading corrections in comparison with those shown on the Fig. \ref{3}. The situation in this case is similar to the one in the CPP.

Furthermore, the calculation of the one loop secular contribution of the third kind to the propagators (in the limit $p\sqrt{\eta_1 \, \eta_2} \to 0$ and $\eta_1/\eta_2 = const$) which follows from the diagrams of the form shown on the Fig. \ref{1}, is also not much different from the dS invariant case. Namely the retarded and advanced propagators again do not receive growing correction in such a limit. At the same time the Keldysh propagator receives correction of the form (\ref{EPP}) with (see \cite{Akhmedov:2013vka} for the details):

\begin{gather}\label{npn0}
n^1_p(\eta) \propto \lambda^2 \, \int \limits_\eta^{{\rm min}(\eta_0, \mu/p)} \frac{du}{u} \, \int_0^\infty \frac{dv}{v} \, \int dl \, l^{D-2}\, \left[\left|A_+\right|^2 \, v^{i\mu} + \left|A_-\right|^2 \, v^{-i\mu}\right] \,h^2\left(l/\sqrt{v}\right)\, \left[h^*\left(l\sqrt{v}\right)\right]^2 + \dots.
\end{gather}
This expression is obtained under the assumption that $n^0_p \gg n^0_q$ for $q \gg p$ and we extend the limits of integration over $l$ and $v$, because these integrals are rapidly converging. The ellipses in (\ref{npn0}) stand for other terms that also describe the change of the level population and vanish when $n_p^0 = 0$. Essentially the RHS of this expression is an analog of the collision integral in Boltzman's kinetic equation \cite{Akhmedov:2013vka}. In the following it is sufficient to realise that $n_p^1(\eta) \propto \lambda^2 \log\left(\eta_0/\eta\right) \, I\left(n^0_p\right)$, where $I\left(n^0_p\right)$ is some kind of the collision integral evaluated for the initial density $n_p^0$.
Similar contribution one obtains for $\kappa_p^0$ and its complex conjugate.

Thus, we obtain that

\begin{gather}\label{np1}
n^1_p(\eta) \propto \left\{\begin{matrix}
\lambda^2 \log \frac{\eta}{\eta_0}, \quad p \ll \frac{\mu}{\eta_0},\\
\lambda^2 \log \frac{p\eta}{\mu},\quad p \gg \frac{\mu}{\eta_0}
\end{matrix}\right.
\end{gather}
In the second loop from the diagram of the Fig. \ref{2} instead of (\ref{n2}) we obtain:

\bqa\label{n22}
n^2_p(\eta) \propto \lambda^4 \, \int \limits^{{\rm min}(\eta_0, \mu/p)}_\eta \frac{du}{u}\, \int_0^\infty \frac{dv}{v} \, \int_0^{\infty} dl \, l^{D-2} \, \left[\left|A_+\right|^2 \, v^{i\mu} + \left|A_-\right|^2 \, v^{-i\mu}\right]\, h^2\left(l/\sqrt{v}\right)\, \left[h^*\left(l\sqrt{v}\right)\right]^2 \times \nn \\ 
\times\int\limits^{\min\left(\eta_0,\frac{\mu u}{l}\right)}_u \frac{du'}{u'}\int_0^\infty \frac{dv'}{v'} \, \int dl' \, \left(l'\right)^{D-2} \, \left[\left|A_+\right|^2 \, \left(v'\right)^{i\mu} + \left|A_-\right|^2 \, \left(v'\right)^{-i\mu}\right]\, h^2\left(l'/\sqrt{v'}\right)\, \left[h^*\left(l'\sqrt{v'}\right)\right]^2 + \dots.
\eqa
The $dl$ integral here can be separated into two regions: $l<\frac{\mu u}{\eta_0}$ and $l>\frac{\mu u}{\eta_0}$. The second region contributes an expression behaving similarly to the eq. \eqref{n2} in the limit $u\to 0$. As was shown above, it does not give an additional power of the logarithm. 

Then we have to estimate only the contribution coming from the region $l< \frac{\mu u}{\eta_0}$:

\bqa\label{n23}
n^2_p(\eta) \propto \lambda^4 \, \int \limits^{{\rm min}(\eta_0, \mu/p)}_\eta \frac{du}{u} \log\left(\frac{\eta_0}{u}\right)\, \int_0^\infty \frac{dv}{v} \, \int_0^{\frac{\mu u}{\eta_0}} dl \, l^{D-2} \, \left[\left|A_+\right|^2 \, v^{i\mu} + \left|A_-\right|^2 \, v^{-i\mu}\right]\, \left[h\left(l/\sqrt{v}\right) h^*\left(l\sqrt{v}\right)\right]^2  \nn
\eqa
where the upper limit of integration of the $l$ integral, $\mu u/ \eta_0$, appears because the contribution of the order $\log\frac{\eta}{\eta_0}$ follows only from this region of momenta, as can be seen from eq. (\ref{np1}). When $u \to 0$, the integral over $l$ in the last expression goes to zero. This indicates that the integral has a polynomial behavior and does not provide higher power of the logarithm.

Thus, it is worthwhile to remark that even if one perturbs the initial BD state in the EPP the resummation of the secular effects remains essentially the same linear problem as in the case of the exact BD in the exact EPP.

\subsection{Secular effects in the sandwich space}

To check whether resummation of the secular effect (or divergence) of the third kind is always a linear problem when there is only expansion (but there is no contraction) we continue with the consideration of the so called sandwich space--time proposed in e.g. \cite{Akhmedov:2017dih}:

\begin{equation}
\label{sandwi}
 ds^2 =
  \begin{cases}
   \left(1+\frac{T^2}{\eta^2+\epsilon^2} \right)\left[d\eta^2-d\vec{x}^2 \right], & \text{} \eta \in (-\infty,0] \\
\frac{T^2}{\epsilon^2}\left[d\eta^2-d\vec{x}^2 \right],   & \text{} \eta \in [0,+\infty),
  \end{cases}
  \quad {\rm where} \quad T^2 \gg \epsilon^2
\end{equation}
This metric describes an expansion between two flat Minkowski spaces at $\eta \ll - T$ and $\eta > - \epsilon$. The expansion stage is very similar to the EPP. 

As is discussed in \cite{Akhmedov:2017dih} free modes in this space can be approximately represented as:

\begin{equation}\label{eq:17}
 f_p(\eta) \approx
  \begin{cases}
  \frac{1}{\sqrt{\omega_{in}}}e^{i\omega_{in}\eta}, & \text{} \eta \ll -T \\
|\eta|^{(D-1)/2}\Bigl[A_p \, H^{(1)}_{i\mu}(p|\eta|) + B_p \, H^{(2)}_{i\mu}(p|\eta|) \Bigr],       & \text{} -T \ll \eta \ll -\epsilon \\
\frac{1}{\sqrt{\omega_{out}}}\left(C_p \, e^{i\omega_{out}\eta} + D_p \,  e^{-i\omega_{out}\eta} \right), 		& \text{} \eta \gg -\epsilon
  \end{cases}
\end{equation}
where $\omega_{in}(p)=\sqrt{\vec{p}^2+m^2}$ and $\omega_{out}(p)=\sqrt{\vec{p}^2+m^2\frac{T^2}{\epsilon^2}}$. The complex coefficients $A_p$, $B_p$, $C_p$ and $D_p$ can be fixed from the gluing conditions at $\eta \sim T$ and $\eta \sim \epsilon$.

These modes can be separated into three calsses

\begin{itemize}

    \item High energy quanta, for which $p|\eta| \gg \mu$ for all the expanding region $\eta \in [-T, -\epsilon]$. These modes do not feel any expansion and do not contribute to the secular growth of interest.
    
    \item Intermediate energy quanta, for which $p\epsilon \ll \mu \ll pT$.
    
    \item Low energy quanta, for which $p|\eta| \ll \mu$ for all the expanding region $\eta \in [-T, -\epsilon]$. These are the modes of the main interest for us.

\end{itemize}

As shown in \cite{Akhmedov:2017dih} during the expansion stage the Keldysh propagator for the low energy and intermediate modes receives secular corrections in the limit $\eta_{1,2} > - \epsilon$ and $T/\epsilon \to \infty$. The corrections are as follows:

\begin{gather}
n^1_p \propto \left\{\begin{matrix}
\lambda^2 \log \frac{\epsilon}{T}, \quad {\rm low} \,\, {\rm energy} \,\, {\rm modes},\\
\lambda^2 \log \frac{p\epsilon}{\mu},\quad {\rm intermidiate} \,\, {\rm modes},
\end{matrix}\right.
\end{gather}
and $n_p^1$ is of order zero for high energy modes. Similar situation appears for $\kappa_p^1$ and its complex conjugate. In \cite{Akhmedov:2017dih} $\lambda \phi^4$ theory was considered in 2D, but similar situation appears in $\lambda \phi^3$ theory at one loop and for any $D$.

Hence, the situation for the sandwich space for the low energy modes is similar to the CPP and it is not hard to see that the diagrams from the Fig. \ref{2} and \ref{3} contribute of the same order. 

Interestingly enough, if one excludes either one of the flat space regions of the entire sandwich space--time and keeps the other, the situation with the IR loop corrections becomes similar to the EPP case. Namely, if one considers space that descrives once started from flat space eternal expansion or nucleation of flat space from zero volume (eternal EPP towards the past, but expansion stops at some moment in the future), then the digrams from the Fig. \ref{2} contribute subleading corrections in comparison with those from the Fig. \ref{3}.

\section{Conclusions and acknowledgements}

In conclusion,  one can respect the dS isometry at each loop level only for massive fields in the EPP with initial BD state. In such a case there are secular effects of the second and third kind and they are related to each other via isometry transformations and analytical continuation in the complex plane of the scalar invariant --- the hyperbolic distance.

Moreover, in the dS invariant situation the problem of the resummmation of the leading secular contributions from all loops reduces to a linear integro--differential Dyson--Schwinger equation, because the diagrams in Fig. \ref{2} provide subleading contributions as compared to those in Fig. \ref{3}.

At the same time the dS isometry is necessarily broken by loop IR divergnces for any initial state in the CPP and global dS. In such a case the resummation of the second type secular contributions still remains to be a linear integro--differential Dyson--Schwinger equation. However, the resummation of the leading IR divergences from all loops now amounts to a nonlinear integro--differential Dyson--Schwinger equation because the digrams in Fig. \ref{2} contribute at the same order as those in Fig. \ref{3}.

All the above results have been shown here for the case when $m > (D-1)/2$ in the units of dS radius. With some modifications they are also going to work  when $\frac{\sqrt{3}}{4}(D-1) < m < (D-1)/2$, because in such a case there is no secular growth in the higher point correlation functions  \cite{Akhmedov:2013vka}. In this case one can put the vertices to their tree--level values in the system of Dyson--Schwinger equations for the propagators and vertices. (Hence, one has to deal with only the equations for the two--point functions.) However, when $m \leq \frac{\sqrt{3}}{4}(D-1)$ one has to solve the combined system of Dyson--Schwinger equations for two--point and higher--point functions together \cite{Akhmedov:2013vka}. That question remains for the moment unsolved.

We would like to acknowledge valuable discussions with A.Polyakov, V.Gorbenko and J.Serreau. AET and UM would like to thank Hermann Nicolai and Stefan Theisen for the hospitality at the Albert Einstein Institute, Golm, where the work on this project was completed.
The work of ETA was supported by the state grant Goszadanie 3.9904.2017/BCh and by the grant from the Foundation for the Advancement of Theoretical Physics and Mathematics “BASIS” and by RFBR grant 18-01-00460 А.

\end{document}